\documentstyle[12pt]{article}
\renewcommand{\theequation}{\arabic{section}.\arabic{equation}}

\begin{document}
\pagestyle{empty}
\vspace* {13mm}
\renewcommand{\thefootnote}{\fnsymbol{footnote}}
\begin{center}
   {\bf ON THE TWO-POINT CORRELATION FUNCTIONS FOR THE \newline
$U_q[SU(2)]$ INVARIANT SPIN ONE-HALF HEISENBERG CHAIN \newline
AT ROOTS OF UNITY}
   \\[25MM]
H. Hinrichsen$^1$, P. P. Martin$^2$, V. Rittenberg$^{1,3}$
and M. Scheunert$^1$ \\[7mm]
{\it $^1$ Physikalisches Institut \\ Nussallee 12,
D-53115 Bonn, Germany \\[5mm]
$^2$ Department of Mathematics, City University\\
Northampton Square, London EC1V\,\,0HB, UK\\[5mm]
$^3$ Rockefeller University, New York,\\
10021 NY, USA}
\\[2.2cm]
{\bf Abstract}
\end{center}
\renewcommand{\thefootnote}{\arabic{footnote}}
\addtocounter{footnote}{-1}
\vspace*{2mm}
%
% Abstract

Using $U_q[SU(2)]$ tensor calculus we compute the two-point
scalar operators (TPSO), their averages on the ground-state
give the two-point correlation functions. The TPSOs are
identified as elements of the Temperley-Lieb algebra and a
recurrence relation is given for them. We have not tempted
to derive the analytic expressions for the correlation functions in
the general case but got some partial results. For $q=e^{i \pi/3}$,
all correlation functions are (trivially) zero, for $q=e^{i \pi/4}$,
they are related in the continuum to the correlation functions of
left-handed and right-handed Majorana fields in the half plane
coupled by the boundary condition. In the case $q=e^{i \pi/6}$,
one gets the correlation functions of Mittag's and Stephen's
parafermions for the three-state Potts model. A diagrammatic
approach to compute correlation functions is also presented.
\vspace{1cm}
\begin{flushleft}
   BONN HE-93-35 \\
   hep-th/9310119 \\
   October 1993 \\
\end{flushleft}
\thispagestyle{empty}
\mbox{}
\newpage
\setcounter{page}{1}
\pagestyle{plain}
\def\leer{\vspace{5mm}}
%
%
%
%%%%%%%%%%%%%%%%%%%%%%%%%%%%%%%%%%%%%%%%%%%%%%%%%%%%%%%%%%%
%
\section{Introduction}
\setcounter{equation}{0}

This paper is a continuation of the effort to compute the
correlation functions for quantum chains invariant under
quantum group transformations with $q$ being a root of
unity. The simplest case which is the $U_q[SU(1/1)]$
invariant chain was considered in Ref. \cite{HR}. Here
we consider the $U_q[SU(2)]$ invariant chain \cite{PS}
described by the Hamiltonian
\begin{equation}
\label{eq11}
H \;=\; -\sum_{j=1}^{L-1}\,e_j
\end{equation}
\noindent
where
\begin{equation}
\label{eq12}
e_j \;=\; -\frac12\,\Bigl(
\sigma_j^x\sigma_{j+1}^x \;+\;
\sigma^y_j\sigma_{j+1}^y \;+\;
\frac{q+q^{-1}}2\,(\sigma^z_j\sigma^z_{j+1}-1) \;+\;
\frac{q-q^{-1}}2\,(\sigma^z_j-\sigma^z_{j+1}) \Bigr)\,.
\end{equation}
In this expression $\sigma^x$,
$\sigma^y$, and $\sigma^z$ are Pauli matrices and
$q$ is a complex parameter.
We consider the antiferromagnetic case. The Hamiltonian
(\ref{eq11}) is the $XXZ$ Heisenberg model with boundary terms.
As is well known, if $q$ is real, one is in the massive phase.
For this case the correlation functions have been recently computed
\cite{DAV} taking first the limit $L \rightarrow \infty$ and
observing that the symmetry is larger than $U_q[SU(2)]$
(the boundary terms disappear). This procedure does not apply
when $q$ is on the unit circle where one is in the massless phase.
If however one takes periodic boundary conditions
(the boundary terms again drop out and the
$U_q[SU(2)]$ symmetry is lost) the correlation functions can be
computed as shown in Ref. \cite{KOR}. We will insist in
maintaining the $U_q[SU(2)]$ symmetry, thus  keeping the boundary
terms. The interest in the calculation of the correlation functions
in this case stems from the observation \cite{ABGR,PS} that
the spectrum of the Hamiltonian (\ref{eq11}) for $q=e^\frac{i \pi}
{m+1}$ contains the spectrum of Hamiltonians corresponding to
statistical models which in the continuum limits are
described by a conformal field theory with a central charge
\begin{equation}
\label{eq13}
c \;=\; 1 - \frac{6}{m(m+1)}\;.
\end{equation}
\indent
In order to explain the problem, let us first consider the case
$q=1$ ($SU(2)$ symmetry). We consider a two-point $SU(2)$ scalar
operator (TPSO)
\begin{equation}
\label{eq14}
g_{l,m} \;=\; -\frac12 \, \vec{\sigma}_l
\cdot \vec{\sigma}_m
\end{equation}
\noindent
and the two-point correlation function
\begin{equation}
\label{eq15}
G(l,m,L) \;=\; \langle 0| g_{l,m} |0\rangle
\;=\; \langle g_{l,m} \rangle\;.
\end{equation}
\noindent
The reader should keep in mind that
$\vec{\sigma}_l$ and $\vec{\sigma}_m$ are
$SU(2)$ tensor operators corresponding to the adjoint
representation and their scalar product is thus
an $SU(2)$ invariant. The ground state is an $SU(2)$ scalar
(we always take an even number of sites). Because of
the boundary terms, translation invariance is lost and the
correlation function is not a function of $(l-m)$ and $L$ only.
The continuum limit is obtained \cite{CA} taking
$L \rightarrow \infty$ with $l$ and $m$ fixed.
\\
\indent
An important observation which is going to be
extensively used in the text is that the
$e_j$ of Eq. (\ref{eq12}) are the generators
of the Temperley-Lieb algebra (with different quotients
when $q^p=\pm 1$) and that the elements
\begin{equation}
\label{eq16}
c_{l,m} \;=\; \frac12 + g_{l,m}\,,
\hspace{2cm}
c_{l,l+1} \;=\; e_l\,
\end{equation}
\noindent
belong to that algebra. In fact, for $l<m<n$ (and still with
$q=1$) these elements satisfy the obvious recurrence relation
\begin{equation}
\label{eq17}
c_{l,n} \;=\; c_{l,m}+c_{m,n}-c_{l,m}\,c_{m,n}-c_{m,n}\,c_{l,m}\,.
\end{equation}
\noindent
It is well known that, considered as elements of the
Temperley-Lieb algebra, the $SU(2)$ scalar states,
of which the ground state is one, build an irreducible
representation of the algebra.
Eq. (\ref{eq17}) thus allows to do the whole calculation of the
correlation functions by purely algebraic means.
\\
\indent
Obviously Eqs. (1.4,6,7) have to be extended for $q \neq 1$.
As will be seen, the two-point scalar operator
(TPSO) $g_{l,m}$ generalizes in two ways and,
unlike the $q=1$ case, $g_{l,m} \neq g_{m,l}$.
One obvious question one would like to answer for
$q=e^\frac{i \pi}{m+1}$ when the continuum theory is
conformal invariant, is what are the surface and bulk exponents
one derives. The question is very relevant since, as we will see,
unlike the $\vec{\sigma}_l$ of the $q=1$ case, the tensor
operators are highly non-local objects and we would like to
know first, whether one can define on the lattice the
proper "local" operators directly for $g_{l,m}$ and get their
critical dimensions.
\\
\indent
The paper is organized as follows. In Sec. 2 we remind the
reader of the definition of tensor operators for the
$U_q[SU(2)]$ quantum group
% \cite{BID,...,SCH}
[7-11]
and give the expression of the
TPSOs which generalize Eq. (\ref{eq14}) for $q \neq 1$.
In Sec. 3 we make the contact with the Temperley-Lieb algebra
(generalizing Eq. (\ref{eq16}) and extending
the recurrence relation
(\ref{eq17}) for $q \neq 1$). At this point, using the Bethe-Ansatz
calculations of Refs.
% \cite{CHIC--DV}
[12-15]
for the ground state
wave function, one could start to compute the correlation functions,
but we did not try to do it. We have preferred to do some exploratory
work which could clarify what to expect from the results to come.
Actually, in Appendix A we give a diagrammatic approach for
the calculation of the correlation functions. This approach
is entirely based on using irreducible representations of
Temperley-Lieb algebras and their quotients. This approach
can be useful for numerical calculations and to obtain
some partial results. One can for example trivially
show that for $q=e^{i \pi/3}$, all the correlation functions
are zero. Another important result for $q=e^{i \pi/4}$
(which corresponds to the Ising model) is that the continuum limit
($L \rightarrow \infty$)  has to be taken with care since one
gets
\begin{equation}
\label{eq18}
\langle g_{2l,2m} \rangle \;=\;
\langle g_{2l+1,2m+1} \rangle \;=\; 0\,,
\end{equation}
\noindent
while the other correlation functions are different from zero.
\\
\indent
In Sec. 4 we compute directly the correlation functions for
$q=e^{i \pi/4}$. This is done using the Ising representation
of the Temperley-Lieb algebra and observing that the
$g_{l,m}$ can be written locally in terms of free fermions.
The correlation functions one obtains are new. In Sec. 5,
still for $q=e^{i \pi/4}$, we compute the time-dependent
correlation functions. The idea is quite simple. Some
time ago, Symanzik \cite{SY} has considered a left-mover and
a right-mover Majorana field in the half-plane
(they are coupled through the boundary conditions)
and computed their propagators. We
perform a conformal transformation from the half-plane
to the strip and discover that the equal-time correlation
functions thus obtained are those derived in Sec. 4.
In Sec. 6 we discuss the physical interpretation of the
correlation functions for $q=e^{i \pi/6}$ which
corresponds to the three-states Potts model.
One discovers that
the "local" operators appearing here in the TPSOs are
the parafermions of Mittag and Stephen~\cite{MS}.
The conclusions are presented in Sec. 7.
%
%
%%%%%%%%%%%%%%%%%%%%%%%%%%%%%%%%%%%%%%%%%%%%%%%%%%%%%%%%%%%
%
\section{Two-point $U_q[SU(2)]$ invariant operators}
\setcounter{equation}{0}
\indent
Let us first again consider the $q=1$ case. The TPSOs given by
Eq. (\ref{eq14}) can be written as follows
\begin{equation}
\label{eq21}
g_{l,m} \;=\; t_l^1t_m^{-1} + t_l^{-1}t_m^1 - t_l^0t_m^0\,,
\end{equation}
\noindent
where
\begin{equation}
\label{eq22}
t_l^{\pm 1} = \mp \sigma^\pm_l\,,\hspace{1cm}
t_l^0=\frac{1}{\sqrt{2}} \sigma_l^z\,. \hspace{1cm}
(\sigma_l^\pm = \frac12 (\sigma_l^x \pm i\sigma_l^y))
\end{equation}
\noindent
The $t_l^p$ ($p=\pm 1,0$) are irreducible tensor
operators corresponding to
the three-dimensional vector representation \cite{WY}
\begin{eqnarray}
\label{eq23}
(ad\,S^\pm)\,t_l^{\pm} &=& 0  \nonumber \\
(ad\,S^\pm)\,t_l^{0} &=& \sqrt{2}\;t_l^{\pm} \\
(ad\,S^\pm)\,t_l^{\mp} &=& \sqrt{2}\;t_l^{0} \nonumber \\
(ad\,S^0)\,t_l^{p} &=& p \;t_l^{p} \nonumber
\hspace{25mm} (p=\pm 1,0)
\end{eqnarray}
\noindent
where
\begin{equation}
\label{eq24}
(ad\,S^r)\,t_l^p \;=\; [S^r,\,t_l^p]
\end{equation}
\noindent
and
\begin{equation}
\label{eq25}
S^\pm \;=\; \sum_{l=1}^L\,\sigma_l^\pm\,,
\hspace{2cm}
S^0 \;=\; \frac12\,\sum_{l=1}^L \, \sigma_l^z
\end{equation}
\noindent
are the $SU(2)$ generators. The TPSOs are obtained from the
tensor operators using Clebsch-Gordan coefficients.
\\
\indent
We now consider the case $q \neq 1$. First let us introduce some
notations:
\begin{eqnarray}
\label{eq26}
S^0_{i,j} &=& \left\{
    \begin{array}{lcc}
    \frac12 \,\sum_{k=i}^j\,\sigma_k^z & \;\; & i\leq j \\
    0 && i>j
    \end{array}
\right\} \\
\label{eq27}
K^\pm_{i,j} &=& q^{\pm S^0_{i,j}} \\
\label{eq28}
S^\pm_{i,j} &=& \sum_{k=i}^j\,K^+_{i,k-1}
\,\sigma_k^\pm \, K^-_{k+1,j}  \hspace{1.5cm} (i<j)\\
\label{eq29}
S^\pm &=& S^\pm_{1,L}\,, \hspace{1cm} K^\pm \;=\; K^\pm_{1,L}\,.
\end{eqnarray}
\noindent
$S^\pm$ and $K^\pm$ commute with the Hamiltonian $H$
(Eq. (\ref{eq11})) and are the generators of the
$U_q[SU(2)]$ algebra:
\begin{eqnarray}
\label{eq210}
&& K^+K^- \;=\; K^-K^+ \;=\; 1 \\
&& K^+S^\pm K^- \;=\; q^{\pm 1} S^\pm \nonumber \\
&& [S^+,S^-] \;=\; \frac{K^{+2}-K^{-2}}{q-q^{-1}}\,. \nonumber
\end{eqnarray}
\indent
Following Ref. \cite{VM} the tensor operators corresponding to the
adjoint representation of $U_q[SU(2)]$ are defined through
the relations
\begin{eqnarray}
\label{eq211}
(ad\,S^\pm)\,t_j^{\pm 1} &=& 0  \\
(ad\,S^\pm)\,t_j^{0} &=& \sqrt{[2]_q}\;t_j^{\pm 1} \nonumber \\
(ad\,S^\pm)\,t_j^{\mp 1} &=& \sqrt{[2]_q}\;t_j^{0} \nonumber \\
(ad\,K^\pm)\,t_j^{p} &=& q^{\pm p}\;t_j^{p} \nonumber \,,
\hspace{25mm} (p=\pm 1,0)
\end{eqnarray}
\noindent
where
\begin{eqnarray}
\label{eq212}
(ad\,S^\pm)\,t_j^p &=& S^\pm\,t_j^p\,K^+ \;-\;
q^{\mp 1}\,K^+\,t_j^p\,S^\pm \\
(ad\,K^\pm)\,t_j^p &=& K^\pm\,t_j^p\,K^\mp \nonumber
\end{eqnarray}
\noindent
and
\begin{equation}
\label{eq213}
[r]_q\;=\;\frac{q^r-q^{-r}}{q-q^{-1}}\,.
\end{equation}
\noindent
These relations generalize Eqs. (\ref{eq23}) and (\ref{eq24}) for
$q \neq 1$. Unlike the $q=1$ case, for $q \neq 1$ \cite{RS}
there are two different sets of tensor operators that we write as
$t_l^p$ and $\tilde{t}_l^p$, their expressions are
\begin{eqnarray}
\label{eq214}
t_j^{-1} &=& \Bigl( K^-_{1,j-1} \Bigr)^2 \,
\tau_j^- \\
\label{eq215}
t_j^{+1} &=& \Bigl( K^+_{1,j-1} \Bigr)^2 \,
\tau_j^+ \;+\;
(q-q^{-1})q^{-1} \,\{\, (q+q^{-1})^{1/2}
K^+_{1,j-1} \, S^+_{1,j-1} \, \tau_j^0 \\
&& \hspace{55mm} +\; (q-q^{-1})(S^+_{1,j-1})^2\,\tau_j^-\, \}
\nonumber \\
\label{eq216}
t_j^0 \hspace{2.2mm}
&=& \tau_j^0\;+\;(q-q^{-1})\,(q+q^{-1})^{1/2}\,
K^-_{1,j-1} S^+_{1,j-1} \, \tau_j^-
\end{eqnarray}
\noindent
respectively
\begin{eqnarray}
\label{eq217}
\tilde{t}_j^{+1} &=& \Bigl( K^-_{1,j-1} \Bigr)^2 \,
\tau_j^+ \\
\label{eq218}
\tilde{t}_j^{-1} &=& \Bigl( K^+_{1,j-1} \Bigr)^2 \,
\tau_j^-\;-\;
(q-q^{-1})q \;\{\, (q+q^{-1})^{1/2}
K^+_{1,j-1} \, S^-_{1,j-1} \, \tau_j^0 \\
&& \hspace{52mm} -\; (q-q^{-1})(S^-_{1,j-1})^2\,\tau_j^+ \,\}
\nonumber \\
\label{eq219}
\tilde{t}_j^{\,0} \hspace{2mm}
&=& \tau_j^0\;-\;(q-q^{-1})\,(q+q^{-1})^{1/2}\,
K^-_{1,j-1} S^-_{1,j-1} \, \tau_j^+\,,
\end{eqnarray}
\noindent
where
\begin{eqnarray}
\label{eq220}
\tau^{\pm 1} &=& \mp q^{\mp 1/2} \, \sigma^\pm \\
\label{eq221}
\tau^0 \hspace{2.8mm}
&=& (q+q^{-1})^{-1/2} \,
\Bigl( \frac12(q-q^{-1}) + \frac12(q+q^{-1})\sigma^z
\Bigr)
\end{eqnarray}
\noindent
are the tensor operators for one site. As shown in Ref.
\cite{SCH}, the tensor operators satisfy the following
permutation properties among the lattice sites:
\begin{eqnarray}
\label{eq222}
t_j^p\,t_i^r &=& \sum_{u,v}\, \hat{R}_{u,v;p,r}\,t_i^u\,t_j^v
\hspace{2cm} (i<j) \\
\label{eq223}
\tilde{t}_i^p\,\tilde{t}_j^r &=&
\sum_{u,v}\, \hat{R}_{u,v;p,r}\,\tilde{t}_j^u\,\tilde{t}_i^v
\hspace{2cm} (i<j) \,,
\end{eqnarray}
\noindent
where the spin $1$ $\otimes$ spin $1\;$ $\hat{R}$
matrix has the following
expression:
\begin{eqnarray}
\label{eqm24}
\hat{R} &=& \sum \hat{R}_{u,v;p,r}\,E^{u,p} \otimes E^{v,r} \nonumber \\
&=& (1-q^{-2}) \,[2]_q\, (q\,E^{1,1} \otimes E^{0,0} +
    q\, E^{0,0} \otimes E^{-1,-1}
+ E^{0,1} \otimes E^{0,-1} \nonumber \\
&&\hspace{2.5cm} + E^{1,0} \otimes E^{-1,0}
+ q(1-q^{-2})\,E^{1,1}\otimes E^{-1,-1}) \nonumber \\
&& + q^2\,(E^{1,1} \otimes E^{1,1} + E^{-1,-1} \otimes E^{-1,-1}) \\
&& + q^{-2}\,(E^{-1,1} \otimes E^{1,-1} + E^{1,-1} \otimes E^{-1,1})
\nonumber \\
&& + E^{0,1} \otimes E^{1,0} + E^{0,-1} \otimes E^{-1,0} +
     E^{1,0} \otimes E^{0,1} + E^{-1,0} \otimes E^{0,-1} +
     E^{0,0} \otimes E^{0,0} \nonumber\,.
\end{eqnarray}
\noindent
Here $E^{u,p}$ is the $3 \times 3$ matrix whose sole
non-vanishing matrix element is on the $u$ th row
and the $p$ th coloumn, and this element is equal to one.
The matrix $\check{R}$ has been calculated from the universal
$R$ matrix of $U_q[SU(2)]$ as given in Ref.~\cite{RM}. Up to
an obvious change of basis in complex $3$-space it
coincides with the $\hat{R}$ matrix of the quantum group
$SO(3)$ \cite{FRT}. Relations like Eqs. (\ref{eq222}),(\ref{eq223}) are well
known
in conformal field theory (see Ref. \cite{MO} and references therein).
\\
\indent
Using $U_q[SU(2)]$ Clebsch-Gordan coefficients we now define
two TPSOs:
\begin{eqnarray}
\label{eq225}
g_{l,m}^+ &=& q^{-2} \, (q^{-1}\,t_l^{+1}\,t_m^{-1}
\;+\; q\, t_l^{-1} \,t_m^{+1} \;-\; t_l^0\,t_m^0 \,) \\
\label{eq226}
g_{l,m}^- &=& q^{+2} \, (q^{-1}\,\tilde{t}_l^{+1}
\,\tilde{t}_m^{-1}  \;+\; q\,
\tilde{t}_l^{-1} \,\tilde{t}_m^{+1} \;-\;
\tilde{t}_l^0\,\tilde{t}_m^0\,)\,.
\end{eqnarray}
\noindent
{}From Eqs. (\ref{eq222}-\ref{eq226}) we derive the following
important relation
\begin{equation}
\label{eq227}
g_{l,m}^\pm \;=\; q^{\pm 4}\,g^\pm_{m,l}\,.
\hspace{2cm}
(l<m)
\end{equation}
\indent
We now give the explicit expressions of the TPSOs.
Using Eqs. (\ref{eq214}-\ref{eq221})
we obtain for $l<m$
\begin{eqnarray}
\label{eq229}
g_{l,m}^+ &=& -\Big[\,
\sigma_l^+ \, (K_{l+1,m-1}^-)^2\, \sigma_m^- \;+\;
\sigma_l^-\,  (K_{l+1,m-1}^+)^2\, \sigma_m^+ \\
&& \;+\; \frac{q+q^{-1}}4 \,(\sigma_l^z \sigma_m^z) \;-\;
\frac14\frac{(q-q^{-1})^2}{(q+q^{-1})} \nonumber \\
&&\;+\; \frac{q-q^{-1}}4 \,(\sigma^z_l-\sigma^z_m)
\nonumber\\
&&\;-\; \frac{q^{-1/2}(q-q^{-1})^2}{2}\,
(S^+_{l+1,m-1}\,K^-_{l+1,m-1}\,\sigma_m^-)
\nonumber\\
&&\;+\; \frac{q^{-1/2}(q^2-q^{-2})}2 \,
(\sigma^z_l\, S_{l+1,m-1}^+\, K_{l+1,m-1}^- \,\sigma^-_m)
\nonumber\\
&&\;-\; \frac{q^{1/2}(q-q^{-1})^2}2 \,
(\sigma^-_l\, S_{l+1,m-1}^+\, K_{l+1,m-1}^+)
\nonumber\\
&&\;-\; \frac{q^{1/2}(q^2-q^{-2})}2\,(
\sigma_l^-\,S_{l+1,m-1}^+ \, K_{l+1,m-1}^+ \, \sigma^z_m)
\nonumber\\
&&\;-\; (q-q^{-1})^2\, (\sigma^-_l\,(S_{l+1,m-1}^+)^2\,
\sigma_m^-)\,\Big]\, \nonumber
\end{eqnarray}
\noindent
and
\begin{eqnarray}
\label{eq230}
g_{l,m}^- &=& -\Big[\,
\sigma_l^+ \, (K_{l+1,m-1}^+)^2\, \sigma_m^- \;+\;
\sigma_l^-\,  (K_{l+1,m-1}^-)^2\, \sigma_m^+ \\
&& \;+\; \frac{q+q^{-1}}4 \,(\sigma_l^z \sigma_m^z) \;-\;
\frac14\frac{(q-q^{-1})^2}{(q+q^{-1})} \nonumber \\
&&\;+\; \frac{q-q^{-1}}4 \,(\sigma^z_l-\sigma^z_m)
\nonumber\\
&&\;-\; \frac{q^{-1/2}(q-q^{-1})^2}{2}\,
(K^-_{l+1,m-1}\,S^-_{l+1,m-1}\,\sigma_m^+)
\nonumber\\
&&\;+\; \frac{q^{-1/2}(q^2-q^{-2})}2 \,
(\sigma^z_l\, K_{l+1,m-1}^- \,S_{l+1,m-1}^-\,\sigma^+_m)
\nonumber\\
&&\;-\; \frac{q^{1/2}(q-q^{-1})^2}2 \,
(\sigma^+_l\, K_{l+1,m-1}^+ \, S_{l+1,m-1}^-)
\nonumber\\
&&\;-\; \frac{q^{1/2}(q^2-q^{-2})}2\,(
\sigma_l^+\, K_{l+1,m-1}^+ \,S_{l+1,m-1}^-\,\sigma^z_m)
\nonumber\\
&&\;-\; (q-q^{-1})^2\, (\sigma^+_l\,(S_{l+1,m-1}^-)^2\,
\sigma_m^+)\,\Big]\,. \nonumber
\end{eqnarray}
\noindent
Notice that
\begin{equation}
\label{eq231}
g_{l,l+1}^+ \;=\; g^-_{l,l+1}\;=\;
e_l-(q+q^{-1})^{-1}\,.
\end{equation}
\noindent
The $g^\pm_{l,m}$ given by Eqs. (\ref{eq229}-\ref{eq230}) generalize
Eq. (\ref{eq14}). The generalization of Eqs. (\ref{eq16}) and
(\ref{eq17}) is going to be presented in the next section.
%
%
%
%%%%%%%%%%%%%%%%%%%%%%%%%%%%%%%%%%%%%%%%%%%%%%%%%%%%%%%%%%%%%%%
%
%
\section{Identification of the TPSOs as elements of
the Temperley Lieb algebra. Recurrence relations.}
\setcounter{equation}{0}
\indent
The Temperley-Lieb algebra \cite{TL} is defined by the
generators $e_j$ ($j=1,\ldots,L-1$) satisfying the
conditions
\begin{eqnarray}
\label{eq31}
&&e_j^2 \;=\; (q+q^{-1})\,e_j \;=\; x \, e_j\\
\label{eq32}
&&e_j e_{j \pm 1} e_j \;=\; e_j \\
\label{eq33}
&&[e_i,e_j]\;=\;0\,. \hspace{2cm} (j \neq i\pm 1)
\end{eqnarray}
\noindent
As already mentioned earlier, the $e_j$ of Eq. (\ref{eq12})
verify the conditions (\ref{eq31}-\ref{eq33}). We now generalize
Eq. (\ref{eq16}) defining
\begin{equation}
\label{eq34}
c_{l,m}^\pm \;=\; g_{l,m}^\pm + x^{-1}
\hspace{2cm} (l<m)
\end{equation}
\noindent
where the TPSOs are defined in (\ref{eq229}-\ref{eq230}).
Here $x=q+q^{-1}$.
The $c^\pm_{l,m}$ are elements of
the Temperley-Lieb algebra.
The proof is simple. We first notice that
\begin{equation}
\label{eq35}
c^\pm_{j,j+1} \;=\; e_j\,.
\end{equation}
\noindent
Moreover, we can show that for $l<m<n$ one has the following
recurrence relation which generalizes Eq. (\ref{eq17}):
\begin{equation}
\label{eq36}
c^\pm_{l,n} \;=\; c_{l,m}^\pm \,+\, c_{m,n}^\pm \,-\,
q^{\pm1}\, c_{l,m}^\pm c_{m,n}^\pm \,-\,
q^{\mp 1}\,c_{m,n}^\pm c_{l,m}^\pm
\end{equation}
\noindent
and this proves our claim. Let us also notice that the $c_{l,m}^\pm$
satisfy the relations
\begin{equation}
\label{eq37}
(c_{l,m}^\pm)^2 \;=\; x\,c_{l,m}^\pm
\end{equation}
\noindent
and for $l<m<n$ with $\mu,\nu=\pm1$
\begin{eqnarray}
\label{eq38}
c_{l,m}^\mu\, c_{m,n}^\nu\, c_{l,m}^\mu &=& c_{l,m}^\mu \\
\label{eq39}
c_{m,n}^\mu\, c_{l,m}^\nu\, c_{m,n}^\mu &=& c_{m,n}^\mu
\end{eqnarray}
\noindent
and for $i<j<l<m$
\begin{equation}
\label{eq310}
[c_{i,j}^\mu,c_{l,m}^\nu ] \;=\;
[c_{i,m}^\mu,c_{j,l}^\nu ] \;=\; 0
\end{equation}
which resemble the relations (\ref{eq31}-\ref{eq33}) for the
generators.
\\
\indent
The relations (\ref{eq36})-(\ref{eq310}) have first been
conjectured on the basis of a detailed computer investigation
of the explicit expressions  (\ref{eq229}), (\ref{eq230}) for the
$g_{l,m}^\pm$. This induced us to study the tensor product of
tensor operators from a more fundamental point of view
\cite{SCH}, which, in particular, led us to the following
factorized expressions for the operators $c^\pm_{l,m}$.
Let $\hat{R}$ be the well-known spin $\frac12$ $\otimes$
spin $\frac12$  $\hat{R}$~matrix (not to be confused with the
matrix in Eq. (\ref{eqm24})) and let us define the operators
\begin{equation}
b_j \;=\; \hat{R}_{j,j+1} \;. \hspace{2cm}
(j=1,2,\ldots,L-1)
\end{equation}
\noindent
Assuming that $\hat{R}$ is normalized appropriately,
we have
\begin{equation}
\label{eq313}
b_j\;=\;q-e_j\,,\hspace{1.5cm}
b^{-1}_j \;=\; q^{-1}-e_j\,,
\end{equation}
\noindent
and then the $c_{l,m}^\pm$ are given by
\begin{equation}
\label{NochEine}
c_{l,m}^\pm \;=\; b_{m-1}^{\pm 1}\, b_{m-2}^{\pm 1} \ldots
b_{l+1}^{\pm 1}\,\,e_l\,\,b_{l+1}^{\mp 1} \ldots
b_{m-2}^{\mp 1} \, b_{m-1}^{\mp 1}
\end{equation}
\noindent
if $l<m$. The aforementioned statements and relations now
follow immediately. Obviously, the relation (\ref{NochEine})
has a wider range of applications than the special case
of the $XXZ$ chain considered here.
For more details we refer the reader
to a forthcoming publication \cite{RS}.
\\
\indent
A direct application of the recurrence relations
(\ref{eq36}) to a diagrammatic approach for the calculation of
the correlation functions is given in Appendix A. Other
applications are shown in Secs. 4 and 6.
\section{Correlation functions for $q=e^{i \pi/4}$}
\setcounter{equation}{0}
\indent
In this section we specialize to the case $q=e^{i \pi/4}$
($x=\sqrt{2}$). The quotient of the Temperley-Lieb algebra for
this case is given in Appendix A but we will not need it here
since we are going to do the calculations in a different way.
We take the Hamiltonian (\ref{eq11}) with an even number of sites
$L=2N$ such that the ground state is an $U_q[SU(2)]$ scalar.
For $x=\sqrt{2}$, the Temperley-Lieb algebra has the
following well-known representation in terms of Pauli matrices
\begin{eqnarray}
\label{eq41}
&& e_{2j} \;=\; c^\pm_{2j,2j+1} \;=\; \frac{1}{\sqrt{2}}\,
(1+\sigma^x_j\sigma^x_{j+1}) \\
\label{eq42}
&& e_{2j-1} \;=\; c^\pm_{2j-1,2j} \;=\; \frac{1}{\sqrt{2}}\,
(1+\sigma^z_j)
\end{eqnarray}
\noindent
which gives the Ising Hamiltonian with free boundary
conditions:
\begin{equation}
\label{eq43}
H \;=\; -\sum_{j=1}^{2N-1}\,e_j \;=\;
-\frac{1}{\sqrt{2}} \, \Bigl(
\sum_{j=1}^{N-1} \, \sigma_j^x \sigma_{j+1}^x \;+\;
\sum_{j=1}^N \, \sigma_j^z \Bigr) \;-\;
\frac{2N-1}{\sqrt{2}}\,.
\end{equation}
\noindent
We use the recurrence relations (\ref{eq36}) to get
\begin{eqnarray}
\label{eq44}
g_{2j,2k}^\pm &=& \pm \frac {i}{\sqrt{2}}\,\tau_j^y \tau_k^y \\
\label{eq45}
g_{2j-1,2k-1}^\pm &=& \pm \frac{i}{\sqrt{2}} \,\tau_j^x \tau_k^x \\
\label{eq46}
g_{2j,2k-1}^\pm &=& - \frac{i}{\sqrt{2}} \,\tau_j^y \tau_k^x \\
\label{eq47}
g_{2j-1,2k}^\pm &=& - \frac{i}{\sqrt{2}} \,\tau_j^x \tau_k^y \,,
\end{eqnarray}
\noindent
where
\begin{equation}
\label{eq48}
\tau_j^x \;=\; \Big( \prod_{i=1}^{j-1} \sigma_i^z \Bigr)
\sigma_j^x\,,
\hspace{2.5cm}
\tau_j^y \;=\; \Big( \prod_{i=1}^{j-1} \sigma_i^z \Bigr)
\sigma_j^y
\end{equation}
\noindent
are fermionic operators:
\begin{eqnarray}
\label{eq49}
\{\tau^x_i,\tau^x_j\} &=& 2\,\delta_{i,j}\,,
\hspace{1cm}
\{\tau^y_i,\tau^y_j\} \;=\; 2\,\delta_{i,j} \\
\{\tau^x_i,\tau^y_j\} &=& 0\,.
\end{eqnarray}
\noindent
Notice that the TPSOs are hermitean operators and that
\begin{equation}
\label{eq410}
g^\pm_{l,m} \;=\; -g^\pm_{m,l}
\end{equation}
\noindent
in agreement with Eq. (\ref{eq227}).
\\
\indent
The two point correlation functions can now be computed
using standard techniques \cite{LSM}. One obtains
\begin{equation}
\label{eq411}
\langle g_{2j,2k}^\pm \rangle \;=\;
\langle g_{2j-1,2k-1}^\pm \rangle \;\,=\;\, 0 \\
\end{equation}
\noindent
in agreement with the results of Appendix A. One also
obtains
\begin{eqnarray}
\label{exact}
\langle g_{2j,2k-1}^\pm \rangle &=&
-\frac{2\sqrt{2}}{2L+1} \, \sum_{n=0}^{L-1}\,
\sin\Bigl(\pi\frac{2n+1}{2L+1}j \Bigr)\,
\cos\Bigl(\pi\frac{2n+1}{2L+1}(k-\frac12) \Bigr)
\\
\langle g_{2j-1,2k}^\pm \rangle &=& \;\;\;
\frac{2\sqrt{2}}{2L+1} \, \sum_{n=0}^{L-1}\,
\sin\Bigl(\pi\frac{2n+1}{2L+1}k \Bigr)\,
\cos\Bigl(\pi\frac{2n+1}{2L+1}(j-\frac12) \Bigr)
\,.
\end{eqnarray}
\noindent
In the
continuum limit ($L \rightarrow \infty$) these
correlation functions reduce to
\begin{eqnarray}
\label{eq412}
\langle g_{2j,2k-1}^\pm \rangle &=& \frac{1}{\pi \sqrt{2}}\,
\Bigl( \frac{1}{R}+\frac{1}{S} \Bigr) \\
\label{eq413}
\langle g_{2j-1,2k}^\pm \rangle &=& \frac{1}{\pi \sqrt{2}}\,
\Bigl( \frac{1}{R}-\frac{1}{S} \Bigr)\,,
\end{eqnarray}
\noindent
where
\begin{equation}
\label{eq414}
R=k-j\,,\hspace{2cm} S=k+j\,.
\end{equation}
\indent
We can now use \cite{CA} the standard definition of
critical exponents: The correlation function is of
the form
\begin{equation}
\label{eq415}
G(R,S) \;=\; \frac{F(\rho)}{R^{2x}}\,,
\hspace{1.5cm}
(\rho = \frac{S^2}{R^2})\,
\end{equation}
\noindent
where $x$ is the bulk critical exponent. From Eqs. (\ref{eq412})
and (\ref{eq413}) we get $x=\frac12$ for both correlation functions.
If
\begin{equation}
\label{eq416}
F(\rho) \;=\; F(1+\alpha)\,;
\hspace{2cm}
\lim_{\alpha \rightarrow 0} \, F(\rho) \;=\; \alpha^{x_s-x}
\end{equation}
\noindent
one gets $x_s=\frac12$ for one correlation function and
$x_s=\frac32$ for the other.
%
%
%%%%%%%%%%%%%%%%%%%%%%%%%%%%%%%%%%%%%%%%%%%%%%%%%%%%%%%%%%%%
%
%
\section{Time-dependent correlation functions for
$q=e^{i \pi/4}$}
\setcounter{equation}{0}
\indent
Our aim is to compute (see Eqs. (\ref{eq44})-(\ref{eq47}))
\begin{eqnarray}
\label{eq51}
\langle \tau_j^y(t_1)\,\tau_k^y(t_2) \rangle &=&
\langle 0|\tau^y(t_1,\theta_1)\,\tau^y(t_2,\theta_2) |0\rangle
\\
\langle \tau_j^x(t_1)\,\tau_k^x(t_2) \rangle &=&
\langle 0|\tau^x(t_1,\theta_1)\,\tau^x(t_2,\theta_2) |0\rangle
\nonumber \\
\langle \tau_j^y(t_1)\,\tau_k^x(t_2) \rangle &=&
\langle 0|\tau^y(t_1,\theta_1)\,\tau^x(t_2,\theta_2) |0\rangle
\nonumber \\
\langle \tau_j^x(t_1)\,\tau_k^y(t_2) \rangle &=&
\langle 0|\tau^x(t_1,\theta_1)\,\tau^y(t_2,\theta_2) |0\rangle
\nonumber
\end{eqnarray}
\noindent
directly in the continuum limit.
Here $\theta_1=ja$, $\theta_2=ka$ and $a$ is the lattice spacing.
We first consider \cite{BIZ}
the Majorana fields $\psi_1(r)$ and $\psi_2(r)$
\begin{equation}
\label{eq52}
\tau^x_j \;=\; \psi_1(r) \,+\, \psi_2(r) \,,
\hspace{2.5cm}
\tau^y_j \;=\; \psi_1(r) \,-\, \psi_2(r)
\end{equation}
\noindent
with the properties:
\begin{equation}
\label{eq53}
\{\psi_\alpha(r_1),\psi_\beta(r_2)\} \;=\; \delta_{\alpha,\beta}
\delta_{r_1,r_2}\,\,,
\hspace{15mm}
\psi_\alpha^\dagger \;=\; \psi_\alpha\,.
\hspace{15mm}
(\alpha,\beta = 1,2)
\end{equation}
\indent
The continuum version of the Hamiltonian (\ref{eq43}) is \cite{BIZ}
\begin{equation}
\label{eq54}
H \;=\; \frac{i}{2}\,
\Bigl(\psi_1(r) \frac{\partial}{\partial r} \psi_1(r) \,-\,
\psi_2(r) \frac{\partial}{\partial r} \psi_2(r)\Bigr)\,.
\end{equation}
\noindent
With
\begin{equation}
\label{eq55}
\psi_\alpha(r,t) \;=\; e^{Ht}\,\psi_\alpha(r)\,e^{-Ht}
\end{equation}
\noindent
we get the equations of motion
\begin{equation}
\label{eq56}
\partial_z\,\psi_1\;=\;0 \,,
\hspace{2cm}
\partial_{\bar{z}}\,\psi_2 \;=\; 0\,,
\end{equation}
\noindent
where $z=r+it$ and $\bar{z}=r-it$. The correlation functions
in the plane are well known:
\begin{eqnarray}
\label{eq57}
\langle \psi_1(z_1)\, \psi_1(z_2) \rangle &=&
\frac{1}{2 \pi (\bar{z}_1-\bar{z}_2)}\,\\
\langle \psi_2(z_1)\, \psi_2(z_2) \rangle &=&
\frac{1}{2 \pi (z_1-z_2)}\,. \nonumber
\end{eqnarray}
\noindent
For our purpose these are not the proper correlation functions.
What we need are the correlation functions in the half plane.
They have been calculated by Symanzik \cite{SY}. The
boundary condition for $\psi_1(r,t)$ and $\psi_2(r,t)$ is
\begin{equation}
\label{eq58}
\tau^x(r,t=0^+) \;=\; \psi_1(r,0^+) + \psi_2(r,0^+) \;=\; 0\,.
\end{equation}
\noindent
This boundary condition couples $\psi_1$ with $\psi_2$
(the other boundary condition $\tau^y(r,t=0^+)=0$ will not be
useful here).
\\
\indent
We define the $2 \times 2$ matrix
\begin{equation}
\label{eq59}
G_{\alpha,\beta} \;=\; \langle \psi_\alpha\,\psi_\beta \rangle
\hspace{2cm}
(\alpha,\beta=1,2)
\end{equation}
\noindent
and the propagators computed in Ref. \cite{SY} are:
\begin{equation}
\label{eq510}
G(z_1,\bar{z}_1,z_2,\bar{z_2}) \;=\; \frac{1}{2\pi} \, \left(
    \begin{array}{cc}
    \frac{1}{\bar{z}_1-\bar{z}_2} &
    \frac{1}{z_1-\bar{z}_2} \\ & \\
    \frac{1}{\bar{z}_1-z_2} & \frac{1}{z_1-z_2}
    \end{array}
\right)\,.
\end{equation}
\noindent
Next we make a conformal transformation which takes the
half-plane into a strip of width~$N$~\cite{CA}
\begin{equation}
\label{eq511}
z \;=\; e^{\frac{\pi}{N}w} \;=\; e^{\frac{\pi}{N}(t+i\theta)}
\end{equation}
\noindent
and get
\begin{equation}
\label{eq512}
G^{strip}(t_1,\theta_1,t_2,\theta_2) \;=\; \frac{1}{2\pi} \, \left(
    \begin{array}{cc}
    \frac{1}{\bar{w}_1-\bar{w}_2} &
    \frac{1}{w_1-\bar{w}_2} \\ & \\
    \frac{1}{\bar{w}_1-w_2} & \frac{1}{w_1-w_2}
    \end{array}
\right)\,.
\end{equation}
\noindent
We denote
\begin{equation}
\label{eq513}
\Delta t = t_2-t_1\,,
\hspace{1cm}
R=\theta_2-\theta_1\,,
\hspace{1cm}
S=\theta_1+\theta_2
\end{equation}
\noindent
and obtain for the correlation functions (\ref{eq51}) the following
expressions:
\begin{eqnarray}
\label{eqa514}
\langle \tau^x(t_1,\theta_1) \,\tau^x(t_2,\theta_2) \rangle &=&
\frac{1}{\pi} \Bigl(\frac{1}{\Delta t+R^2/\Delta t}\,+\,
                     \frac{1}{\Delta t+S^2/\Delta t}
\Bigr) \nonumber \\
\langle \tau^y(t_1,\theta_1) \,\tau^y(t_2,\theta_2) \rangle &=&
\frac{1}{\pi} \Bigl(\frac{1}{\Delta t+R^2/\Delta t}\,-\,
                     \frac{1}{\Delta t+S^2/\Delta t}
\Bigr) \\
\langle i\,\tau^y(t_1,\theta_1) \,\tau^x(t_2,\theta_2) \rangle &=&
-\frac{1}{\pi} \Bigl(\frac{1}{R+\Delta t^2/R}\,+\,
                     \frac{1}{S+\Delta t^2/S}
\Bigr) \nonumber\\
\langle i\,\tau^x(t_1,\theta_1) \,\tau^y(t_2,\theta_2) \rangle &=&
-\frac{1}{\pi} \Bigl(\frac{1}{R+\Delta t^2/R}\,-\,
                     \frac{1}{S+\Delta t^2/S}
\Bigr)
\end{eqnarray}
\noindent
We now get convinced that we have taken the proper boundary condition
(\ref{eq58}) since taking $\Delta t=0$, we recover the
results of Sec. 4 (see Eqs. (\ref{eq411}-\ref{eq413})).
The second boundary condition of Ref. \cite{SY}
$(\tau^y(r,t=0^+)=0)$ is useful if one changes the overall
sign of the Hamiltonian~(\ref{eq54}).
%
%
%%%%%%%%%%%%%%%%%%%%%%%%%%%%%%%%%%%%%%%%%%%%%%%%%%%%%%%%%%%
%
%
\section{Physical interpretation of the correlation functions
for $q=e^{i \pi/6}$}
\setcounter{equation}{0}
\indent
We saw in the last two sections that for $q=e^{i \pi/4}
\hspace{2mm}$ ($g^\pm_{l,m}=-g^\pm_{m,l}$),
the correlation functions were
given by fermionic propagators in the strip. In the present case,
we have
\begin{equation}
\label{eq61}
g^\pm_{l,m} \;=\; \omega^{\pm 1} \, g_{m,l}\,.
\hspace{2cm}
(l<m,\;\; \omega = e^{2 \pi i/3})
\end{equation}
\noindent
This suggests that the "local" operators in the subspace where
the irreducible representation of the Temperley-Lieb
algebra acts are the parafermions of Mittag and Stephen \cite{MS}
occurring in the three-state Potts model.
\\
\indent
We first give the Temperley-Lieb generators for the $q=e^{i \pi/6}$ quotient:
\begin{eqnarray}
\label{eq62}
e_{2j} &=& \frac{1}{\sqrt{3}} \,
(1\,+\, \Gamma_j \Gamma_{j+1}^\dagger \,+\,
\Gamma_j^\dagger \Gamma_{j+1} ) \\
e_{2j-1} &=& \frac{1}{\sqrt{3}} \,
(1\,+\, \sigma_j + \sigma_j^\dagger )\,, \nonumber
\end{eqnarray}
where
\begin{equation}
\label{eq63a}
\sigma\Gamma \;=\; \omega\,\Gamma\sigma\,,
\hspace{2cm}
\sigma^3\;=\; \Gamma^3\;=\; 1
\end{equation}
\noindent
with the representation:
\begin{equation}
\label{eq63b}
\sigma\;=\;
\left(
    \begin{array}{ccc}
    1 && \\
    & \omega & \\
    && \omega^2
    \end{array}
\right)\,,
\hspace{2cm}
\Gamma\;=\;
\left(
    \begin{array}{ccc}
    && 1 \\
    1 && \\
    & 1 &
    \end{array}
\right)\,.
\end{equation}
\noindent
Using Eqs (\ref{eq11}) and (\ref{eq62}) one obtains
\begin{equation}
\label{eq64}
H \;=\; -\sum_{j=1}^{2N-1}\,e_j \;=\;
-\frac{1}{\sqrt{3}} \, \Bigl(
\sum_{j=1}^{N-1} \,
(\Gamma_j \Gamma_{j+1}^\dagger+\Gamma_j^\dagger \Gamma_{j+1})
\;+\; \sum_{j=1}^N \,
(\sigma_j+\sigma_j^\dagger) \Bigr) \;-\;
\frac{2N-1}{\sqrt{3}}\;.
\end{equation}
\indent
We now define the parafermionic operators \cite{MS}
\begin{eqnarray}
\label{eq65}
\Pi_j &=& \Bigl( \prod_{i=1}^{j-1} \sigma_i \Bigr) \, \Gamma_j \\
Q_j &=& \Gamma_j^\dagger \, \Bigl( \prod_{i=1}^{j}
\sigma_i^\dagger \Bigr)  	\nonumber
\end{eqnarray}
\noindent
with
\begin{equation}
\label{eq66}
\Pi_j^2 \;=\; \Pi^\dagger_j\,,
\hspace{2.5cm}
Q_j^2 \;=\; Q^\dagger_j
\end{equation}
\noindent
and (for $l<m$)
\begin{eqnarray}
\label{eq67}
Q_m\,Q_l &=& \omega\,Q_l\,Q_m  \\
Q_m\,\Pi_l &=& \omega^2\,\Pi_l\,Q_m \nonumber \\
\Pi_m\,Q_l &=& \omega^2\,Q_l\,\Pi_m \nonumber \\
\Pi_m\,\Pi_l &=& \omega\,\Pi_l\,\Pi_m \nonumber \,.
\end{eqnarray}
\noindent
Beside $Q_l$ and $\Pi_l$ it is also useful to consider their
complex conjugates $Q_l^*$ and $\Pi_l^*$. In terms of
parafermions the ferromagnetic three-state Potts
Hamiltonian is:
\begin{eqnarray}
\label{eq68}
H &=& - \Bigl[ \, \sum_{j=1}^{N-1} \, (Q_j \Pi_{j+1} \,+\,
\Pi^\dagger_{j+1} Q_{j}^\dagger) \;+\;
\sum_{j=1}^N \, (\Pi_j Q_j \,+\,
Q_j^\dagger \Pi_j^\dagger) \;+\; \frac{2N+1}{\sqrt{3}} \Bigr]
\\
&=& - \Bigl[ \, \sum_{j=1}^{N-1} \, (Q_j^* \Pi_{j+1}^* \,+\,
\Pi^{*\dagger}_{j+1} Q_{j}^{*\dagger}) \;+\;
\sum_{j=1}^N \, (\Pi^*_j Q^*_j \,+\,
Q_j^{*\dagger} \Pi_j^{*\dagger}) \;+\;
\frac{2N+1}{\sqrt{3}} \Bigr]\,.
\nonumber
\end{eqnarray}
\noindent
Like in the Ising case, we take $L=2N$, where $L$ is the number
of sites in the $XXZ$ chain. We use again the recurrence relations
(\ref{eq36}) and notice that we can express the
$g^\pm_{l,m}$ in terms of parafermions. With $l<m$, we have:
\begin{eqnarray}
\label{eq610}
g^+_{2l,2m} \hspace{7mm} &=& \frac{1}{\sqrt{3}}\,
(\omega Q_l Q_m^\dagger \,+\, \omega^2 Q_m Q_l^\dagger)
\\
g^+_{2l,2m-1} \hspace{4mm} &=& \frac{1}{\sqrt{3}}\,
(Q_l \Pi_m \,+\, \Pi_m^\dagger Q_l^\dagger)
\nonumber \\
g^+_{2l-1,2m} \hspace{4mm} &=& \frac{1}{\sqrt{3}}\,
(\Pi_l  Q_m \,+\, Q_m^\dagger  \Pi_l^\dagger)
\nonumber \\
g^+_{2l-1,2m-1} &=& \frac{1}{\sqrt{3}}\,
(\omega \Pi_l \Pi_m^\dagger \,+\, \omega^2 \Pi_m \Pi_l^\dagger)
\nonumber
\end{eqnarray}
\noindent
Notice that the $g^+_{l,m}$ are hermitean. We also obtain
\begin{equation}
\label{eq69b}
g_{l,m}^- \;=\; (g^+_{l,m})^* \,.
\end{equation}
\noindent
Since the Hamiltonian  and the ground state
are invariant under conjugation and
so is the ground state, the correlation functions one
obtains from $g^+_{l,m}$ and $g^-_{l,m}$ are the same
\begin{equation}
\langle g^+_{l,m} \rangle \;=\;  \langle g^-_{l,m} \rangle
\end{equation}
\noindent
as one can check from the examples computed in Appendix A
(see Eqs. (\ref{eqa17})-(\ref{eqa17b})).
%
%
%%%%%%%%%%%%%%%%%%%%%%%%%%%%%%%%%%%%%%%%%%%%%%%%%%%%%%%%%%%%%%%%%%%%
%
%
\section{Conclusions}
\setcounter{equation}{0}
We think that we have clarified some properties of the
$U_q[SU(2)]$ invariant two-point correlation functions
for $q^p=\pm1$. From the examples we considered we have seen
that although  the objects one has to compute
(see Eqs. (\ref{eq229}) and (\ref{eq230})) look
hopeless, by taking proper sequences of sites
(see Eqs. (\ref{eq411})-(\ref{eq413})) one can discover
new local operators with a clear physical meaning
(see Eqs. (\ref{eq44})-(\ref{eq47}) and (\ref{eq610}))
in terms of which the correlation functions are
expressed in a transparent way. The key points
are the recurrence and the symmetry relations
(\ref{eq36}) and (\ref{eq227}) which help formulating
the whole problem as a purely algebraic one.
The same picture should be valid for the correlation functions
for other quantum chains where the underlying
algebra is not anymore the Temperley-Lieb
one (see Eq. (\ref{NochEine}) which is more general). As we
have often done in the text, we would like refer the
reader to the Appendix A where a diagrammatic approach
to the calculation of correlation functions is given.
The full range of applications of this method is not yet
known to us.
As the reader might have guessed, this paper will have
an obvious sequel.
\\[5mm]
{\bf Acknowledgements.}
We would like to thank Rainald Flume for pointing out
to us Ref.~\cite{SY}. We would also like to thank
Alexander Berkovich for a careful reading of the manuscript.
%
%
%%%%%%%%%%%%%%%% Diagram Definitions %%%%%%%%%%%%%%%%%%%%%%%%%%
%
%
\def\lu{\begin{picture}(1,2)\put(1,0){\oval(2,4)[bl]}\end{picture}}
\def\lo{\begin{picture}(1,2)\put(1,0){\oval(2,4)[tl]}\end{picture}}
\def\ru{\begin{picture}(1,2)\put(-1,0){\oval(2,4)[br]}\end{picture}}
\def\ro{\begin{picture}(1,2)\put(-1,0){\oval(2,4)[tr]}\end{picture}}
\def\hu{\begin{picture}(2,2)\put(1,0){\oval(2,4)[b]}\end{picture}}
\def\ho{\begin{picture}(2,2)\put(1,0){\oval(2,4)[t]}\end{picture}}
\def\li{\begin{picture}(1,5)\put(0,0){\line(0,1){5}}\end{picture}}
\def\lh{\begin{picture}(4,1)\put(0,0){\line(1,0){4}}\end{picture}}
\def\lli{\begin{picture}(1,6)\put(0,0){\line(0,1){6}}\end{picture}}
\def\lkv{\begin{picture}(1,2)\put(0,0){\line(0,1){2}}\end{picture}}
\def\lkh{\begin{picture}(2,1)\put(0,0){\line(1,0){2}}\end{picture}}
\def\lkkv{\begin{picture}(1,1)\put(0,0){\line(0,1){1}}\end{picture}}
\def\lkkh{\begin{picture}(1,1)\put(0,0){\line(1,0){1}}\end{picture}}
\def\fo{\footnotesize}
\def\nu{\begin{picture}(7,2)(0.4,0.6)
                         \put(0,0){\fo 1}\put(2,0){\fo 2}
                         \put(4,0){\fo 3}\put(6,0){\fo 4}
                         \end{picture}}
\def\nus{\begin{picture}(7,2)(0.5,0.55)
                         \put(0,0){\fo 1'}\put(2,0){\fo 2'}
                         \put(4,0){\fo 3'}\put(6,0){\fo 4'}
                         \end{picture}}

\def\da{\begin{picture}(8,4.5)(0,4.5)
			\put(1,7){\hu} \put(1,2){\ho}
                        \put(5,2){\li} \put(7,2){\li}
			\put(1,1){\nus} \put(1,8){\nu}
                        \end{picture}\vspace{10mm}}
\def\db{\begin{picture}(8,4.5)(0,4.5)
			\put(3,7){\hu} \put(3,2){\ho}
                        \put(1,2){\li} \put(7,2){\li}
			\put(1,1){\nus} \put(1,8){\nu}
                        \end{picture}}
\def\dc{\begin{picture}(8,4.5)(0,4.5)
			\put(5,7){\hu} \put(5,2){\ho}
                        \put(1,2){\li} \put(3,2){\li}
			\put(1,1){\nus} \put(1,8){\nu}
                        \end{picture}}
\def\dd{\begin{picture}(8,7)(0,7)
			\put(1,12.3){\hu} \put(3,1.7){\ho}
			\put(3,6.7){\hu} \put(1,7.3){\ho}
                        \put(1,1.7){\li} \put(7,1.7){\li}
			\put(7,7.3){\li} \put(5,7.3){\li}
			\put(1,0.7){\nus} \put(1,13.3){\nu}
                        \end{picture}\vspace{15mm}}
\def\de{\begin{picture}(8,6)(0,6)
			\put(1,10){\hu} \put(3,2){\ho}
                        \put(7,2){\li} \put(7,5){\li}
			\put(1,2){\lkv} \put(1,4){\lo}
			\put(2,6){\lkh} \put(5,8){\ru}
			\put(5,8){\lkv}
			\put(1,1){\nus} \put(1,11){\nu}
                        \end{picture}}
\def\df{\begin{picture}(8,7.5)(0,7.5)
			\put(1,13){\hu} \put(1,2){\ho}
			\put(1,7){\hu} \put(3,7){\ho}
                        \put(5,2){\li} \put(7,2){\li}
			\put(7,7){\li} \put(7,12){\lkkv}
			\put(5,12){\lkkv} \put(1,8){\lo}
			\put(2,10){\lkh} \put(5,12){\ru}
			\put(1,7){\lkkv}
			\put(1,1){\nus} \put(1,14){\nu}
                        \end{picture}\vspace{7mm}}
\def\dg{\begin{picture}(8,7)(0,7)
			\put(1,12){\hu} \put(1,7){\ho}
			\put(1,7){\hu} \put(1,2){\ho}
                        \put(5,2){\li} \put(5,7){\li}
                        \put(7,2){\li} \put(7,7){\li}
			\put(1,1){\nus} \put(1,13){\nu}
                        \end{picture}\vspace{5mm}}
\def\dh{\begin{picture}(8,4)(0,4)
			\put(3,7){\hu} \put(3,1){\ho}
                        \put(1,1){\lli} \put(7,1){\lli}
                        \end{picture}\vspace{9mm}}
\def\di{\begin{picture}(8,4)(0,4)
			\put(5,7){\hu} \put(5,1){\ho}
                        \put(1,1){\lli} \put(3,1){\lli}
                        \end{picture}}
\def\dj{\begin{picture}(8,4)(0,4)
			\put(3,7){\hu} \put(5,1){\ho}
                        \put(1,1){\lli} \put(3,1){\lkkv}
			\put(3,2){\lo} \put(4,4){\lkh}
			\put(7,6){\ru} \put(7,6){\lkkv}
                        \end{picture}}
\def\dk{\begin{picture}(8,4)(0,4)
			\put(5,7){\hu} \put(3,1){\ho}
                        \put(1,1){\lli} \put(7,1){\lkkv}
			\put(7,2){\ro} \put(4,4){\lkh}
			\put(3,6){\lu} \put(3,6){\lkkv}
                        \end{picture}}
\def\dm{\begin{picture}(8,4)(0,4)
			\put(3,7){\hu} \put(1,1){\ho}
                        \put(5,1){\ho} \put(1,6){\lkkv}
			\put(1,6){\lu} \put(7,6){\ru}
			\put(2,4){\lh} \put(7,6){\lkkv}
                        \end{picture}}
\def\ds{\begin{picture}(8,4)(0,4)
			\put(3,1){\ho} \put(1,7){\hu}
                        \put(5,7){\hu} \put(1,1){\lkkv}
			\put(1,2){\lo} \put(7,2){\ro}
			\put(2,4){\lh} \put(7,1){\lkkv}
                        \end{picture}}
\def\dl{\begin{picture}(8,4)(0,4)
			\put(1,1){\ho} \put(5,1){\ho}
                        \put(1,7){\hu} \put(5,7){\hu}
                        \end{picture}\vspace{8mm}}
\def\dq{\begin{picture}(12,4.5)(0,4.5)
\put(1,6){\lkv} \put(1,6){\lu} \put(2,4){\lh} \put(6,4){\lh}
\put(11,6){\ru} \put(11,6){\lkv}
\put(3,7){\lkkv} \put(3,7){\lu} \put(4,5){\lh}
\put(9,7){\ru} \put(9,7){\lkkv} \put(5,8){\hu}
\put(1,1){\ho} \put(5,1){\ho} \put(9,1){\ho}
\end{picture}}

\def\dpp{\begin{picture}(12,4.5)(0,4.5)
\put(1,7){\lkkv} \put(1,7){\lu} \put(2,5){\lh}
\put(7,7){\ru} \put(7,7){\lkkv}
\put(3,8){\hu} \put(9,8){\hu}
\put(1,1){\ho} \put(5,1){\ho} \put(9,1){\ho}
\end{picture}}

\def\doo{\begin{picture}(12,4.5)(0,4.5)
\put(5,7){\lkkv} \put(5,7){\lu} \put(6,5){\lh}
\put(11,7){\ru} \put(11,7){\lkkv}
\put(1,8){\hu} \put(7,8){\hu}
\put(1,1){\ho} \put(5,1){\ho} \put(9,1){\ho}
\end{picture}}

\def\dr{\begin{picture}(12,4.5)(0,4.5)
\put(1,7){\lkkv} \put(1,7){\lu} \put(2,5){\lh}
\put(6,5){\lh} \put(11,7){\ru} \put(11,7){\lkkv}
\put(3,8){\hu} \put(7,8){\hu}
\put(1,1){\ho} \put(5,1){\ho} \put(9,1){\ho}
\end{picture}\vspace{2cm}}

\def\dn{\begin{picture}(12,4.5)(0,4.5)
\put(1,8){\hu} \put(5,8){\hu} \put(9,8){\hu}
\put(1,1){\ho} \put(5,1){\ho} \put(9,1){\ho}
\end{picture}\vspace{10mm}}

\def\dt{\begin{picture}(6,4)(0,4)
\put(1,7){\hu} \put(3,1){\ho} \put(1,1){\lkkv}
\put(1,2){\lo} \put(2,4){\lkh} \put(5,6){\ru} \put(5,6){\lkkv}
\end{picture}\vspace{9mm}}

\def\du{\begin{picture}(6,4)(0,4)
\put(1,1){\ho} \put(3,7){\hu} \put(1,6){\lkkv}
\put(1,6){\lu} \put(2,4){\lkh} \put(5,2){\ro} \put(5,1){\lkkv}
\end{picture}}

\def\dw{\begin{picture}(10,5)(0,5)
\put(1,9){\hu} \put(1,1){\lkv} \put(1,3){\lkkv}
\put(1,4){\lo} \put(2,6){\lkh}
\put(5,8){\ru} \put(5,8){\lkkv}
\put(3,1){\lkv} \put(3,3){\lo} \put(4,5){\lkh}
\put(7,7){\ru} \put(7,7){\lkv}
\put(5,1){\lkkv} \put(5,2){\lo} \put(6,4){\lkh}
\put(9,6){\lkkv} \put(9,7){\lkv} \put(9,6){\ru}
\put(7,1){\ho}
\end{picture}\vspace{11mm}}

\def\dx{\begin{picture}(10,5)(0,5)
\put(1,1){\ho} \put(5,1){\ho}
\put(3,9){\hu} \put(7,9){\hu}
\put(1,7){\lkv} \put(1,7){\lu}
\put(2,5){\lh} \put(6,5){\lkh}
\put(9,3){\ro} \put(9,1){\lkv}
\end{picture}}

\def\dy{\begin{picture}(10,5)(0,5)
\put(1,1){\ho} \put(7,1){\ho} \put(3,9){\hu}
\put(1,8){\lkkv} \put(1,8){\lu} \put(2,6){\lh}
\put(7,8){\ru} \put(7,8){\lkkv}
\put(5,1){\lkv} \put(5,3){\lo} \put(6,5){\lkh}
\put(9,7){\ru} \put(9,7){\lkv}
\end{picture}}

\def\dz{\begin{picture}(10,5)(0,5)
\put(1,9){\hu} \put(7,9){\hu} \put(5,1){\ho}
\put(1,1){\lkv} \put(1,3){\lkkv} \put(1,4){\lo}
\put(2,6){\lkh} \put(5,8){\ru}
\put(5,8){\lkkv} \put(3,1){\lkkv} \put(3,2){\lo}
\put(4,4){\lh} \put(9,2){\ro} \put(9,1){\lkkv}
\end{picture}}

\def\dma{\begin{picture}(6,4)(0,4)
\put(1,7){\hu} \put(1,1){\ho} \put(5,1){\lli}
\end{picture}\vspace{8mm}}

\def\dmb{\begin{picture}(6,4)(0,4)
\put(3,7){\hu} \put(3,1){\ho} \put(1,1){\lli}
\end{picture}}

\def\dmc{\begin{picture}(6,4)(0,4)
\put(3,7){\hu} \put(1,1){\ho} \put(5,1){\lkkv}
\put(5,2){\ro} \put(2,4){\lkh} \put(1,6){\lu}
\put(1,6){\lkkv}
\end{picture}}

\def\dmd{\begin{picture}(6,4)(0,4)
\put(1,7){\hu} \put(3,1){\ho} \put(1,1){\lkkv}
\put(1,2){\lo} \put(2,4){\lkh} \put(5,6){\ru}
\put(5,6){\lkkv}
\end{picture}}

\def\dme{\begin{picture}(6,4)(0,4)
\put(1,1){\lli} \put(3,1){\lli} \put(5,1){\lli}
\end{picture}}

\def\dmf{\begin{picture}(10,5)(0,5)
\put(1,1){\ho} \put(7,1){\ho}
\put(3,9){\hu} \put(7,9){\hu}
\put(1,7){\lkv} \put(1,7){\lu}
\put(2,5){\lkh}
\put(5,3){\ro} \put(5,1){\lkv}
\end{picture}\vspace{11mm}}

\def\dmg{\begin{picture}(10,5)(0,5)
\put(3,1){\ho} \put(7,1){\ho}
\put(1,9){\hu} \put(7,9){\hu}
\put(1,1){\lkv} \put(1,3){\lo}
\put(2,5){\lkh}
\put(5,7){\ru} \put(5,7){\lkv}
\end{picture}}

\def\dmi{\begin{picture}(10,5)(0,5)
\put(3,1){\ho} \put(7,1){\ho}
\put(1,9){\hu} \put(5,9){\hu}
\put(1,1){\lkv} \put(1,3){\lo}
\put(2,5){\lkh} \put(4,5){\lh}
\put(9,7){\ru} \put(9,7){\lkv}
\end{picture}}

\def\dmk{\begin{picture}(10,5)(0,5)
\put(1,1){\ho} \put(7,1){\ho}
\put(5,9){\hu} \put(1,7){\lkv}
\put(1,7){\lu} \put(2,5){\lkh}
\put(5,3){\ro} \put(5,1){\lkv}
\put(3,8){\lkkv} \put(3,8){\lu}
\put(4,6){\lh} \put(9,8){\ru}
\put(9,8){\lkkv}
\end{picture}}
%
%
%%%%%%%%%%%%%%%%%%%%%%%%%%%%%%%%%%%%%%%%%%%%%%%%%%%%%%%%%%%%%%
%
%
\appendix
\renewcommand{\theequation}
{\Alph{section}.\arabic{equation}}
\setlength{\unitlength}{2.3mm}
\thicklines
\section{A diagrammatic calculation of the
correlation functions}
\setcounter{equation}{0}
\indent
Since at this point we have not yet been able to compute the
two-point correlation functions~$g_{l,m}^\pm$ for $q^p=\pm 1$
except in the case $p=4$, we would like to suggest a diagrammtic
approach~\cite{PPM} to the problem which is complementary to the
usual Bethe-Ansatz method. This different approach cannot only
bring a different insight to the problem of computing correlation
functions but, as will be seen at the end of this appendix, give
unexpected results.
\\
\indent
We start by giving a diagrammatic description to the generators
$c_{j,j+1}$ of the Temperley-Lieb algebra (see Eq. (\ref{eq47})).
All the rules can be understood taking a chain with four sites.
The three generators $c_{1,2}$, $c_{2,3}$ and $c_{3,4}$ are
represented as:
\begin{equation}
\label{eqa1}
c_{1,2}\;=\; \da\,, \hspace{15mm}
c_{2,3}\;=\; \db\,, \hspace{15mm}
c_{3,4}\;=\; \dc\,.
\end{equation}
\noindent
The diagrams should always be followed from the top to the bottom.
The word $c_{1,2}c_{2,3}$ can be obtained connecting the
lower part of $c_{1,2}$ with the upper part of $c_{2,3}$:
\begin{equation}
\label{eqa2}
c_{1,2}\,c_{2,3} \;=\; \dd \;=\; \de \,.
\end{equation}
\noindent
Notice that through multiple applications of $c_{j,j+1}$ one never
generates intersecting lines.
If we now multiply $c_{1,2}c_{2,3}$ with $c_{1,2}$
(put $c_{1,2}$ to the bottom of (\ref{eqa2})) we obtain
\begin{equation}
\label{eqa3}
\df \;=\; c_{1,2}\,c_{2,3}\,c_{1,2} \;=\; c_{1,2} \;=\; \da
\end{equation}
\noindent
in agreement with Eq. (\ref{eq43}). A new rule appears if one
multiplies $c_{1,2}$ with itself:
\begin{equation}
\label{eqa4}
\dg \;=\; x \, \da \;=\; x\,c_{1,2}
\;=\; (q+q^{-1})\,c_{1,2}\;.
\end{equation}
\noindent
Thus a closed blob is taken away from the diagram and is substituted
by the number $x = q+q^{-1}$. The diagrammatic
calculation described by Eq. (\ref{eqa4}) gives the defining identity
(\ref{eq43}). One can now use the recurrence relation (\ref{eq48})
to get the diagrammatic expression of the $c_{l,m}^\pm$.
For example
\begin{equation}
\label{eqa5}
c_{2,4}^+ \;=\; c_{2,3} + c_{3,4} - q\,c_{2,3}\,c_{3,4}
- q^{-1}\,c_{3,4}\,c_{2,3}
\end{equation}
\noindent
has the following diagram
\begin{equation}
\label{eqa5b}
c_{2,4}^+ \;\;=\;\; \dh \;+\; \di \;-\; q\,\dj \;-\; q^{-1}\,\dk
\end{equation}
\indent
We want to do the calcualtion of the correlation functions purely
algebraically. The ground-state of the system is a
$U_q[SU(2)]$ scalar. It has been shown \cite{PPM} that the
linear combination of words
corresponding to the scalars are of a special type.
The lower part is disconnected from the upper part and the
lower part is the same for all the words (it can be taken
in an arbitrary way). For example there are two scalars
appearing in the four sites problem and an eigenword
$w$ can be written as
\begin{equation}
\label{eqa6}
w \;=\; a_1\, \dl \;+\; a_2\,\dm\,,
\end{equation}
\noindent
where the constants $a_1$ and $a_2$ have to be determined from
the eigenvalue problem (each diagram can be understood as a
basis ket vector). Similarly, for the six sites problem (there
are five $U_q[SU(2)]$ scalars in this case) the ground-state
has the following representation:
\begin{eqnarray}
\label{eqa7}
w\;&=&\;\;a_1\,\dn \,+\, a_2\, \doo \,+\, a_3\, \dpp \\[0.9cm]
&&+\, a_4\, \dq \,+\, a_5\, \dr\,. \nonumber \\[3mm]
&& \nonumber
\end{eqnarray}
\indent
In order to find the eigenwords of $H$ given by Eq. (\ref{eq11}),
we solve the equation
\begin{equation}
\label{eqa8}
H\,w \;=\; \lambda \,w\,.
\end{equation}
\noindent
This is easily done for the four-site case. Using
Eqs. (\ref{eqa1}) and
(\ref{eqa6}), one obtains:
\begin{eqnarray}
\label{eqa9}
c_{1,2}\,w &=& (a_1 x+a_2) \,\dl \,\nonumber \\[8mm]
c_{2,3}\,w &=& (a_1+a_2 x) \,\dm \,	\\[8mm]
c_{3,4}\,w &=& (a_1 x +a_2) \,\dl \,, \nonumber
\end{eqnarray}
\noindent
and instead of Eq. (\ref{eqa8}) we get
\begin{equation}
\label{eqa9b}
(2a_1x+2a_2+\lambda a_1) \,\dl \;+\;
(a_1+a_2x+\lambda a_2) \, \dm \;=\; 0
\end{equation}
\noindent
from which we obtain
\begin{eqnarray}
\label{eqa10}
\lambda^{(I)} &=& -\frac{3x+\sqrt{x^2+8}}{2}\,,
\hspace{2cm}
a_1^{(I)} \;=\; \frac{a_2^{(I)}}{2}\,(x+\sqrt{x^2+8}) \\
\lambda^{(II)} &=& -\frac{3x-\sqrt{x^2+8}}{2}\,,
\hspace{2cm}
a_1^{(II)} \;=\; \frac{a_2^{(II)}}{2}\,(x-\sqrt{x^2+8})
\nonumber \,.
\end{eqnarray}
\noindent
Obviously $\lambda^{(I)}$ corresponds to the ground-state energy.
One can check that
\begin{equation}
\label{eqa11}
w^T \;=\; a_1\,\dl \,+\,a_2\,\ds\,,
\end{equation}
\noindent
where $a_1$ and $a_2$ correspond to the two solutions given by
Eq. (\ref{eqa10}), are left eigenwords:
\begin{equation}
\label{eqa12}
w^T\,H \;=\; \lambda \, w^T\,.
\end{equation}
\noindent
One can also check that the normalization condition
\begin{equation}
\label{eqa13}
w^T\,w \;\;=\;\; 1\;\dl
\end{equation}
\noindent
gives
\begin{equation}
\label{eqa14}
(a_1^2+a_2^2)\,x^2 \,+\, 2\,x\,a_1a_2 \;=\; 1
\end{equation}
\noindent
and that one has
\begin{equation}
\label{eqa15}
w^{T(I)}w^{(II)} \;= \; w^{T(II)}w^{(I)} \;=\; 0\,.
\end{equation}
\noindent
We now compute correlation functions taking averages on the word
$w$ given by (\ref{eqa6}) without specifying whether it is the
eigenword corresponding to the ground-state or not:
\begin{equation}
\label{eqa16}
w^T c_{l,m}\,w \;=\; \langle c_{l,m} \rangle \,\dl\,.
\end{equation}
\noindent
This calculation can be done using the diagrammatic approach,
one obtains:
\begin{eqnarray}
\label{eqa17}
\langle c^\pm_{1,2} \rangle &=& \langle c^\pm_{3,4} \rangle
\;=\; x\,(a_1x+a_2)^2 \\
\langle c^\pm_{2,3} \rangle &=&
\langle c^\pm_{1,4} \rangle \;=\;
x \, (a_1+x a_2)^2 \\
\langle c_{1,3}^\pm \rangle &=& \langle
c_{2,4}^\pm \rangle  \;=\;
x \bigl[ a_1^2+a_2^2+a_1a_2\,(3x-x^3) \bigr]\,.
\label{eqa17b}
\end{eqnarray}
\noindent
As the reader has already noticed, the advantage of the diagrammatic
approach is that one avoids to work with matrices altogether
and that one does not have to use $q$-Young symmetrizers~\cite{Q}
in order to construct the $U_q[SU(2)]$ singlet states.
\\
\indent
Let us now return to our results given by Eqs. (\ref{eqa17})-
(\ref{eqa17b})
and make a few observations. We first take $x=1$
$(q=e^{i \pi/3})$. Using the normalization condition
(\ref{eqa14}) one discovers that
\begin{equation}
\label{eqa18}
\langle c^\pm_{l,m} \rangle \;=\; 1
\end{equation}
\noindent
which implies (see Eq. (\ref{eq46})) that
\begin{equation}
\label{eqa19}
\langle g^\pm_{l,m} \rangle \;=\; 0\,.
\end{equation}
\noindent
This result can be understood in the following way. For $x=1$,
the Temperley-Lieb algebra has the quotient \cite{PM}
\begin{equation}
\label{eqa20}
e_j\;=\; 1\,.
\end{equation}
\noindent
This makes (using Eq. (\ref{eq36})) all the $c_{l,m}^\pm=1$ and
$g^\pm_{l,m}=0$ in the general case (not only for four sites).
\\
\indent
Let us next rewrite $\langle g_{1,3}^\pm \rangle$ using the
normalization condition (\ref{eqa14}), we obtain:
\begin{equation}
\label{eqa21}
\langle g_{1,3}^\pm \rangle \;=\; a_1a_2\,
(x^2-1)\, (2-x^2)\,.
\end{equation}
\noindent
One notices that $\langle g_{1,3}^\pm \rangle
= \langle g_{2,4}^\pm \rangle$ vanishes not only for
$x=1$ but also for $x = \sqrt{2}$ ($q=e^{i \pi /4}$).
This result is not new for the average on the ground state
(energy $\lambda^{(I)}$ in Eq. (\ref{eqa10})) as was shown by
explicit calculation in Eq. (\ref{eq411}), but it is new for
the excited state (energy $\lambda^{(II)}$ in Eq. (\ref{eqa10})).
Using the diagrammatic approach, one can show that for
any number of sites and $x=\sqrt{2}$ one has
\begin{equation}
\label{eqa22}
\langle g_{2j,2k}^\pm \rangle \;=\; \langle
g_{2j-1,2k-1}^\pm \rangle \;=\; 0
\end{equation}
\noindent
{\it {for all singlet states}}.
We first sketch the proof through
two examples. For $x=\sqrt{2}$,
the Temperley-Lieb algebra has the quotient \cite{PM}:
\begin{equation}
\label{eqa23}
\sqrt{2}\,(c_{k,k+1}+c_{k+1,k+2})\,-\,c_{k,k+1}\,c_{k+1,k+2}
\,-\, c_{k+1,k+2}\,c_{k,k+1} \;=\; 1\;.
\end{equation}
\noindent
Other quotients corresponding to different values of $x$
are also described in Ref. \cite{PM}. Using Eq. (\ref{eqa23})
we get for example
\begin{equation}
\label{eqa24}
g_{1,3}^\pm \;=\; \mp\frac{i}{\sqrt{2}}\,c_{1,2}c_{2,3}
\,\pm\, \frac{i}{\sqrt{2}}\,c_{2,3}c_{1,2}\;.
\end{equation}
\noindent
Thus $g_{1,3}^\pm$ has the diagram
\begin{equation}
\label{eqa25}
g_{1,3}^\pm \;=\; \mp\frac{i}{\sqrt{2}}\,\dt
\;\pm\, \frac{i}{\sqrt{2}}\,\du\;.
\end{equation}
\noindent
Notice that $(g_{1,3}^\pm)^T=g_{1,3}^\pm$. The transposition
means (see also Eq. (\ref{eqa12})) taking the
reflected (up-down) diagrams.
Since $w$ and $w^T$ in the calculation of averages (see Eq.
(\ref{eqa16})) also correspond to up-down reflected diagrams,
the contributions of the two diagrams in (\ref{eqa25}) cancel.
One can go one step further.
Quite generally, the recurrence relation (\ref{eq36}) is
equivalent to
\begin{equation}
\label{ManfredRecurrence}
g^\pm_{l,m} \;=\; -q^{\pm 1}\,g_{l,m}^\pm\,g_{m,n}^\pm -
q^{\mp 1}\,g_{m,n}^\pm\,g_{l,m}^\pm
\end{equation}
\noindent
for $l<m<n$. In particular, we have
\begin{equation}
\label{man}
g_{1,5}^\pm \;=\; -\,q^{\pm 1} \, g_{1,3}^\pm g_{3,5}^\pm
\,-\,q^{\mp 1} \, g_{3,5}^\pm g_{1,3}^\pm\,.
\end{equation}
\noindent
The diagram corresponding to $g_{1,3}^\pm \,g_{3,5}^\pm$ is
\begin{equation}
\label{eqa27}
-\frac12 \,\left( \dw \,+\, \dx \right) \;+\;
\frac12 \,\left( \dy \,+\, \dz \right) \,.
\end{equation}
\noindent
Explicit calculation of $w^Tg_{1,3}^\pm\,g_{3,5}^\pm$ shows that the
contributions of the two terms in Eq. (\ref{eqa27}) cancel,
and the same is true for $g_{3,5}^\pm\,g_{1,3}^\pm$.
\\
\indent
In order to give a general proof we start from the preceding
example and re-express $g_{1,3}^\pm\,g_{3,5}^\pm$
using the Ising quotient relation (\ref{eqa23}) in the form
\begin{equation}
\dme \;\;=\;\; \sqrt{2}\,\left(
\dma \,+\, \dmb \right) \;-\;
\left(\dmc \,+\, \dmd \right)
\end{equation}
\noindent
to eliminate diagrams with three descending lines. We can get,
for example,
\begin{eqnarray}
g_{1,3}^\pm\,g_{3,5}^\pm
&=&\frac12\Biggl[ \sqrt{2} \left(
\nonumber
\dmf \,-\, \dmg \right) \;+\; \left( \dx\,-\,\dmi \right) \\
&&\hspace{5mm}+\;\; \left(\dz \,-\, \dmk \right) \,\Biggr]
\end{eqnarray}
\noindent
so that
\begin{equation}
\label{eqz2}
g_{1,3}^\pm\,g_{3,5}^\pm \;=\; -(g_{1,3}^\pm\,g_{3,5}^\pm)^T\,.
\end{equation}
\noindent
Since $(g_{1,3}^\pm\,g_{3,5}^\pm)^T =
(g_{3,5}^\pm)^T(g_{1,3}^\pm)^T$ and, for
example, $(g^\pm_{1,3})^T = - g_{1,3}^\pm$
(see Eq. (\ref{eqa25}), we conclude that
\begin{equation}
\label{Mittwoch}
g_{1,3}^\pm\,g_{3,5}^\pm + g_{3,5}^\pm\,g_{1,3}^\pm
\;=\; 0\,.
\end{equation}
\noindent
Let us next show that
\begin{equation}
\label{Marke}
(g_{l,l+2k}^\pm)^T \;=\; -g_{l,l+2k}^\pm
\end{equation}
in the Ising case. In fact, from the recurrence
relation (\ref{ManfredRecurrence}) we note that
$g_{l,n}^\pm$ is effectively linear in $g_{l,l+2}^\pm$.
Consequently, using Eq. (\ref{Mittwoch}) and
the fact that $[g_{i,j}^\pm,\,g_{k,l}^\pm] = 0$ for
$i<j<k<l$, we obtain
\begin{equation}
\label{eqz3}
g_{l-2,l}^\pm \, g_{l,n}^\pm + g_{l,n}^\pm \, g_{l-2,l}^\pm
\;=\; 0\,.
\end{equation}
\noindent
Eq. (\ref{Marke}) now follows by induction on $k$, the case $k=1$
being settled by Eq. (\ref{eqa25}).
\\
\indent
Finally, we are ready to prove Eq. (\ref{eqa22}).
Consider any element $w$ of the Temperley-Lieb algebra
of the form $w=Y e_1e_3\ldots e_{2N-1}$, where $Y$
is an arbitrary element of the algebra (thus $w$
is an element of the left ideal generated
by $e_1e_3\ldots e_{2N-1}$). If $X$ is one more
element of the Temperley-Lieb algebra, we have
\begin{equation}
w^T\,X\,w \;=\; \chi(X)
\,e_1e_3\ldots e_{2N-1}\,,
\end{equation}
\noindent
where $\chi(X)$ is just a complex number
(to see this, draw the corresponding diagram).
Since obviously $(e_1e_3\ldots e_{2N-1})^T=e_1e_3\ldots e_{2N-1}$,
this implies that
\begin{equation}
\label{eqz1}
w^T\,X\,w \;=\; (w^T\,X\,w)^T \;=\; w^T\,X^T\,w\,.
\end{equation}
\noindent
Consequently, this expression vanishes whenever
$X^T=-X$, in particular, for
$X=g_{l,l+2k}^\pm$.
%
%
%%%%%%%%%%%%%%%%%%%%%%%%%%%% REFERENCES %%%%%%%%%%%%%%%%%%%%%%%%%
%
%


\begin{thebibliography}{99}
\bibitem{HR}
    H. Hinrichsen and V. Rittenberg,
    {\it Phys. Lett.} {\bf B 304} (1993) 115
\bibitem{PS}
    V. Pasquier and H. Saleur, {\it Nucl. Phys.} {\bf B 330}
    (1990) 523
\bibitem{DAV}
    H. Jimbo, K. Miki and A. Nakayashiki,
    {\it Phys. Lett.} {\bf A 168} (1992) 256
\bibitem{KOR}
    V. E. Korepin {\it Comm. Math. Phys.} {\bf 86} (1982) 392
\bibitem{ABGR}
    F. C. Alcaraz, M. Baake, U. Grimm and V. Rittenberg,
    {\it J. Phys.} {\bf A 22} (1989) L5
\bibitem{CA} J. L. Cardy, {\it Nucl. Phys.}
   {\bf B 240} (1984) 514
\bibitem{BID}
    L. C. Biedenharn and M. Tarlini, {\it Lett. Math. Phys.}
    {\bf 20} (1990) 271
\bibitem{MACK}
    G. Mack and V. Schomerus, {\it Nucl. Phys.}
    {\bf B 370} (1992) 185
\bibitem{VM}
    V. Rittenberg and M. Scheunert, {\it J. Math. Phys.}
    {\bf 33} (1992) 436
\bibitem{SCH}
    M. Scheunert, The tensor product of tensor operators
    over quantum algebras, Bonn preprint BONN-HE-93-28 (1993)
\bibitem{RS}
    M. Scheunert, to be published
\bibitem{CHIC}
    F. C. Alcaraz, M. N. Barber, M. T. Batchelor, R. J. Baxter
    and G. R. W. Quispel, \\ {\it J. Phys.} {\bf A 20} (1987) 6397
\bibitem{SK}
    E. K. Sklyanin, {\it J. Phys.} {\bf A 21} (1988) 2375
\bibitem{LR}
    L. Mezincescu and R. Nepomechie, {\it Mod. Phys. Lett.}
    {\bf A 6} (1991) 2467
\bibitem{DV}
    C. Destri and H. J. de Vega,
    {\it Nucl. Phys.} {\bf B 374} (1992) 692,
    {\bf B 385} (1992) 361
\bibitem{SY}
    K. Symanzik, {\it Nucl. Phys.} {\bf B 190} (1981) 1
\bibitem{MS}
    L. Mittag and M. J. Stephen, {\it J. Math. Phys.}
    {\bf 12} (1971) 441
\bibitem{WY}
    B. G. Wyborne, "Classical Groups for Physicists",
    Ed. John Wiley, New York 1974
\bibitem{RM}
    A. N. Kirillov and N. Yu. Reshetikhin, Representations
    of the algebra $U_q(su(2))$, $q$-orthogonal
    polynomials, and invariants of links, LOMI preprint
    E-9-88, Leningrad (1988)
\bibitem{FRT}
    N. Yu. Reshetikhin, L. A. Takhtadzhyan and L. D. Faddeev,\\
    {\it Leningrad Math. J.} {\bf 1} (1990) 193
\bibitem{MO}
    G. Moore and N. Reshetikhin, {\it Nucl. Phys.} {\bf B 328}
    (1989) 557
\bibitem{TL}
    H. N. V. Temperley and E. H. Lieb, {\it Proc. R. Soc.}
    {\bf A 322} (1971) 251
\bibitem{LSM}
    E. Lieb, T. Schultz and D. Mattis,
    {\it Ann. Phys.} {\bf 16} (1961) 407
\bibitem{BIZ}
    M. Bander and C. Itzykson, {\it Phys. Rev.}
    {\bf D 15} (1977) 463; \\
    J. B. Zuber and C. Itzykson, {\it Phys. Rev.}
    {\bf D 15} (1977) 2875
\bibitem{PPM}
    P. P. Martin, "Potts Models and Related Problems in Statistical
    Mechanics",\\ World Scientific, Singapore 1991, and references
    included
\bibitem{Q}
    R. Dipper and G. James, {\it Proc. London Math. Soc.}
    {\bf 54} (1987) 57
\bibitem{PM}
    P. P. Martin, {\it J. Phys.} {\bf A 22} (1989) 3103,
    {\bf A 23} (1990) 7
\end{thebibliography}
\end{document}